# Rigidity and flexibility of biological networks

*Merse E. Gáspár and Peter Csermely*

**Abstract**
The network approach became a widely used tool to understand the behaviour of complex systems in the last decade. We start from a short description of structural rigidity theory. A detailed account on the combinatorial rigidity analysis of protein structures, as well as local flexibility measures of proteins and their applications in explaining allostery and thermostability is given. We also briefly discuss the network aspects of cytoskeletal tensegrity. Finally, we show the importance of the balance between functional flexibility and rigidity in protein–protein interaction, metabolic, gene regulatory and neuronal networks. Our summary raises the possibility that the concepts of flexibility and rigidity can be generalized to all networks.

Corresponding author. Peter Csermely, Department of Medical Chemistry, Semmelweis University, Tűzoltó str. 37-47, H-1094 Budapest, Hungary. Tel: 361 459 1500, extn: 60130; FAX: 361 266 3802; E-mail: csermely.peter@med.semmelweis.univ.hu

**Merse E. Gáspár** is a research fellow in the Department of Medical Chemistry, Semmelweis University. He is a physicist, his PhD thesis was written in the field of general relativity. As a member of the LINK-Group, his main interests are in structural rigidity theory and its applications, percolation phenomena, random walks and pebble game on weighted graphs.

**Peter Csermely** is a professor in the Department of Medical Chemistry, Semmelweis University. He established the multi-disciplinary network science LINK-Group (www.linkgroup.hu) in 2004. His main interests are the topology, dynamics and function of complex networks, including their critical transitions and the application of this knowledge to understand aging, disease states and enrich drug design.

# INTRODUCTION

**The network concept**

Various types of networks (represented by weighted, directed, coloured graphs) became widely applied to describe complex systems in the last decade. Modelling a complex system as a network is appropriate and practical, when it is built up from similar, but distinct objects (represented by nodes/vertices of the graph) and decisively pair-wise interactions (represented by edges/links). However, especially in biological systems, where data coverage often has technical difficulties, not all of the possible interactions are strong enough to take them into account. The complex functions are usually related to network structure. Most self-organized networks (and biological networks in particular) are small worlds with scale-free degree distribution [1, 2]. It is important to note, that for the functional analysis of the complex system, network topology is not enough, but network dynamics has to be also included. In dynamic network models quantities assigned to nodes and/or edges may vary, and/or the background topology itself may also change, where the latter phenomenon is commonly referred as "network evolution".

The fact that biological sciences use more and more network models is not accidental. Biological entities can endure evolutionary selection, if they can cooperatively interact [3]. Therefore, networks form a hierarchy from molecular level to societies [4]. In many of these networks, nodes represent individual physical units, such as amino acids of protein structural networks [5–7]. Other networks are abstract conceptualizations, such as the gene regulatory network [8, 9], where nodes represent DNA segments, and their connections are often indirect representing the interactions between the transcriptional processes of the corresponding genes. Definition of the most important biological networks can be found in several papers and books [4, 10, 11]. However, it is important to note that network definition may represent a rather difficult problem itself. The definition of the entities representing nodes, their connections representing edges and especially the weight of the interaction may not be straightforward. In most cases, edge weights refer to the strength or probability of the interaction [12–14]. However, in numerous studies edge weights are neglected.

**Rigidity and flexibility**

Rigidity and flexibility have a lot of meanings in everyday life and in science as well. We can talk about rigidity of materials, frameworks or behaviours. The properties: stability, stiffness, robustness are often used together with the concept of rigidity. In this paper, rigidity and flexibility are used as properties, which characterize the possible states that can be reached by the system under certain type of external influences. This concept implies that flexibility also measures the internal degree of freedom and its distribution in the system. Corresponding to the above definition, two kinds of rigidity/flexibility concepts are used in the literature: 1.) structural and 2.) functional rigidity/flexibility. We talk about the structural rigidity of networks, when the network represents the geometrical structure of a complex system. This means that the network is a framework, which is embedded in a metric space. In simplest cases, this can be studied by the so-called mathematical theory of structural rigidity, which was born with the early paper of Maxwell written on constraint counting in 1864 [15]. This theory is introduced in the next section, followed by its applications studying protein flexibility. Closing the sections on structural rigidity, properties of cytoskeletal network will be reviewed in conjunction with the biotensegrity concept [16], since it plays an important role in the structure and dynamics of the cell.

Contrary to structural rigidity/flexibility, in the case of functional rigidity/flexibility, the network has to be dynamic (or has to undergo evolution), but the network need not to be embedded into metric space. Since both network dynamics and the external influence provoking this dynamics depend on the actual problem studied, there is no general theory of functional rigidity/flexibility. Moreover, introducing dynamics makes the solution both conceptually and computationally more difficult. Therefore, much less studies deal with functional rigidity than those on structural rigidity. It is important to note that the concept of internal degree of freedom is valid for both structural and functional rigidities, since functional rigidity means that the system or living organism has few possible responses to an external influence. For example, if there is no internal degree of freedom, the system (or its behaviour) is said to be totally rigid in both cases. However, complex systems (including biological systems) can be flexible/adaptive and rigid/stable/robust at the same time, usually on different time scales. In neurophysiology, this is called as the stability–flexibility dilemma [17], but achieving the appropriate balance between rigidity and flexibility is a key question for all biological networks at all levels. The most important factor in this problem is the connection between the topological structure and the functional flexibility properties of the system [18]. We will conclude the review by the summary of functional rigidity/flexibility studies for the most common cellular networks and neuronal networks as well as by listing a few major perspectives of the field.

## STRUCTURAL RIGIDITY THEORY IN A NUTSHELL

Structural rigidity theory deals with networks, in which, 1.) nodes are physical objects embedded in a metric space in a natural way, and 2.) links represent geometrical constrains. The simplest model in structural rigidity theory is the bar-joint framework in Euclidean space [19]. The bar-joint framework represents a graph where nodes (joints) are embedded as points (without expansion) in the space, and edges (bars) represent distance constraints, with fixed distance, between the connected joints. If distance constraints are allowed to be inequalities instead of strict equalities, we get the so-called tensegrity generalization. The physical representation of this latter model consists of rods, cables (which cannot be compressed) and struts (which may have negative stress) [20–22].

The framework is said to be rigid, if continuous motions (allowed by the constraints) are trivial congruencies. The dimension of the space is an important parameter for rigidity. N points in the d-dimensional space has $d \cdot N$ degrees of freedom, and the number of isometries is equal to $d \cdot (d+1)/2$, therefore sufficient condition for rigidity in three dimension is that the number of bars $\geq 3 \cdot N - 6$. This is called Maxwell counting rule [15].

Infinitesimal rigidity is a stronger condition, which means that the framework has only trivial infinitesimal flexes [19, 23]. An infinitesimal flex (see small vectors in Figure 1) is a continuous infinitesimal motion of the framework, in which the displacement vectors of the nodes are perpendicular to the rods. Therefore, rods cannot act with force against the infinitesimal displacement. In practice, this means that an infinitesimally rigid framework cannot wobble.

Nontrivial infinitesimal flexes usually exist in case of special geometric situations (for example collinear joints, see the most left framework in Figure 1). A framework is said to be generically rigid, if we can get a rigid framework introducing a minor change to its coordinate values [19, 24]. In this case only the underlying graph topology is important instead of the geometry and the framework lacks any special symmetry. However, rigid systems with special symmetries have an increasing importance e.g. in the field of crystallization [25, 26] (including protein crystals [27]), and in the recently improved theory of infinite frameworks [28]. For further rigidity and stability properties for finite frameworks, see [21, 29].

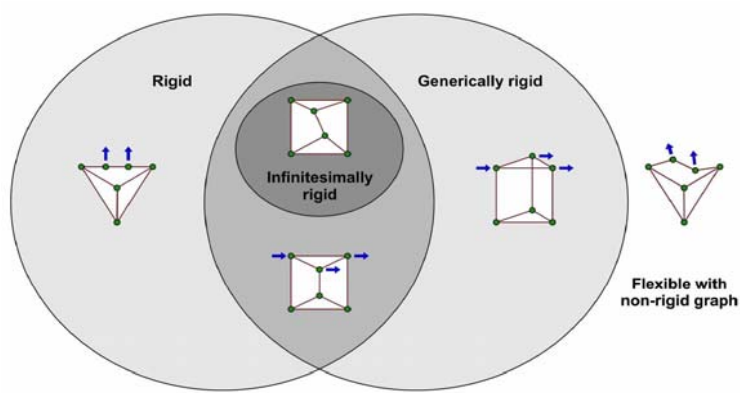

**Figure 1:** Rigidity properties illustrated by 2-dimensional bar-joint frameworks. Arrows show non trivial flex for infinitesimally not rigid examples. Note, that in the generically rigid set each example has the same graph. Since, one of them is infinitesimally rigid, this means that the graph is said to be rigid. A very similar map is presented in [19].

Generic rigidity or rigidity of the graph can be studied by combinatorial graph theory. However, in the last decades it became clear that matroid theory, an axiomatized theory of independence sets, is the best tool studying generic rigidity [30, 31]. This is because only independent constrains may decrease the degrees of freedom of the system. If an interaction is added to an already rigid region of the network (called as redundant edge), the region becomes over-rigid (or stressed), but this does not modify the rigid cluster decomposition of the network, which is the main problem of rigidity theory.

A complete combinatorial characterization of graph rigidity is given by Laman's theorem for two dimensions, which states that the graph is rigid, if and only if, it is (2,3)-critical [32]. However, this result has not been generalized to three dimensions for bar-joint frameworks yet. It is of note that for body-bar-hinge frameworks (where "body" is an extended object with six degree of freedom in three dimensions) the Laman theorem can be generalized [33, 34]. In this case the necessary and sufficient condition of the theorem is that the underlying multi-graph has to be (6,6)-critical, in which a hinge is represented by five multiple edges (loss of five degrees of freedom), since a hinge allows only the rotation around one axis. See Figure 2 for further explanation.

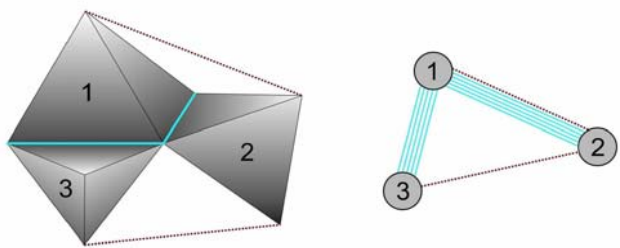

**Figure 2:** Illustration of body-bar-hinge frameworks and their multi-graph representation. Numbered tetrahedrons represent rigid bodies in the body-bar-hinge framework on the left. However rigid bodies may have any kind of shapes. Bars (dashed red lines) can connect any points belonging to rigid bodies, but here vertices of the tetrahedrons are used. Hinges (solid blue lines) are located at common edges of tetrahedrons. We note that, in the generic case, lines belonging to bars and hinges may not be parallel. In the associated multi-graph (right): body, bar and hinge is represented by node, (dashed) edge and five (solid) edges (connecting nodes belong to the interacting bodies) respectively. The above example of a body-bar-hinge framework is minimally rigid (having no redundant bars or hinges) corresponding to the fact that its associated multi-graph is (6,6)-critical, which means that it has exactly $6 \cdot c - 6$ edges, and its sub-graphs have maximum $6 \cdot c' - 6$ edges, where $c$ and $c'$ are the numbers of nodes in the graph and sub-graphs, respectively.

Combinatorial graph theory has a lot of polynomial algorithms [35, 36], but the most widely used algorithm is the so called pebble game algorithm [37, 38] performing rigid cluster decomposition and determining redundant edges and the internal degree of freedom.

## RIGIDITY AND FLEXIBILITY OF PROTEIN STRUCTURES

Rigidity and flexibility of proteins at the right place play an important role in protein function [39–42]. To describe protein rigidity properties mathematical structural rigidity theory is a useful tool, since proteins are real molecular frameworks. However, interactions between atoms cannot be considered as simple geometrical constrains, since they may fluctuate. Thus rigidity analysis of protein structures provides information on the phase space of collective motions of protein segments. However, local flexibility/mobility properties of protein structures at the atomic level are also very important, and require other methods of protein dynamics for their study not reviewed here.

**Structural rigidity analysis of proteins**

Computer programs are available for the rigid cluster decomposition of proteins using three-dimensional protein structures. Figure 3 illustrates the two possible major ways of modelling molecular frameworks on the ethane molecule. The first method uses the bar-joint model, in which, each atom is represented by a joint, and distance constraints of bonds are represented by bars. For modelling angle constraints of rotating bonds, such as single covalent bonds, additional bars connecting next nearest neighbours (shown by dashed lines on panel B of Figure 3) have to be added to the model. Torsional constraints of double covalent (or peptide and resonant) bonds can be fixed by introducing third-nearest distance constraints. The resulting network is the so called bond-bending network used and analysed by MSU-FIRST software [38].

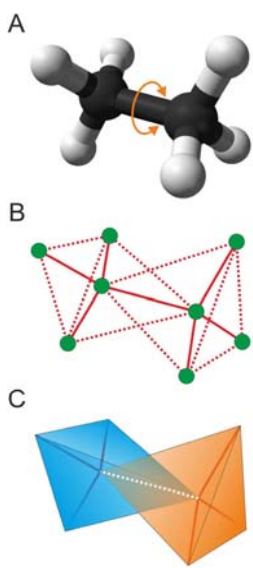

**Figure 3:** (A) Molecular framework models are illustrated with the help of the ethane molecule having a rotating single covalent bond between carbon atoms. (B) In the bond-bending network nodes represent atoms, solid lines represent distance constrains belongs to central forces, dashed lines represent additional distance constraints to model angle constraints of rotating bonds. (C) In the body-bar-hinge representation rigid bodies are set of atoms rigidly attached to each others. Rotational degree of freedom of a single covalent bond is modelled as a hinge, shown by dashed (white) line in the figure. The associated multi-graph (for definition see Figure 2) is used for rigidity analysis.

In the body-bar-hinge model, local rigid groups of atoms are represented by rigid bodies. For example, in the ethane molecule ($C_2H_6$) both carbons with their four bonds form a rigid tetrahedron (see panel C of Figure 3), and the rotating single covalent bond between the carbons is modelled by a hinge. Double covalent bond between carbons does not allow any rotational freedom, therefore an ethylene molecule ($C_2H_4$) itself is a rigid body. Weaker interactions can be modelled by specifying a number of bars (from 1 to 6), each one removing one degree of freedom. The body-bar-hinge model is implemented in by the KINARI web server [43]. The default model of KINARI uses hinges for modelling single covalent-, disulphide- and strong hydrogen bonds, six bars for double covalent- and resonant bonds, and two bars for hydrophobic interactions.

Since a hinge can be modelled by 5 general bars, an equivalent body-bar model (see Figure 4 in [34]) can also be used, which is implemented in the later versions of FIRST (see ASU-FIRST [44]). Both FIRST and KINARI use the pebble game algorithm. It is also possible to set the energy cut-off value for bonds and interactions included to the analysis. For KINARI this selection can be performed separately for different type of bonds. KINARI has the other great advantage, that it has an on-line visualization plug-in for studying the results of the rigidity analysis (http://kinari.cs.umass.edu). Both FIRST and KINARI give the rigid cluster decomposition of the protein and a number of calculated rigidity measures, such as the widely used rigidity order parameter meaning the relative size of the largest rigid cluster as compared to the whole structure. It is also possible to classify each interaction in terms of its contribution to the overall rigidity using redundancy analysis [43].

Rigid cluster decomposition methods have also been applied for other macromolecules, such as for RNA [45–47]. Since molecular dynamics calculations producing individual pathways in protein folding are computationally very expensive, the much faster rigid cluster decomposition methods are used to reduce the conformation space efficiently and improve dynamical simulations [48–52].

Rigid cluster decomposition is a very useful tool, but it is important to note its limitations.
- In both molecular framework models, we have to decide, which bonds and interactions are treated as a strict and exact integer number of constraints, and all other forces are completely ignored.
- To make this decision, energy cut-off values have to be introduced, but there is no clear methodology how to choose adequate cut-off values, as well as compare and interpret the results obtained, since rigidity is a non-local property, and rigid cluster decomposition is very sensitive to the cut-off values [53].
- Generic rigidity depends on the network topology and not on the geometry. In reality the possibility that the framework is not generic is close to zero, but there are always lots of nodes near degenerate position (for example arranged in a straight line), which can be important, since real chemical bonds have dispersion in length and dihedral angles.

The above arguments suggest that for real structures rigidity can be characterised more likely with the help of probabilistic quantities, rather than by definite cluster decomposition. The recently developed generalization of the pebble game algorithm by Gonzalez *et al.* [54] offers a very helpful solution to this problem. This generalization is called as "virtual pebble game", and represents a non-integer probabilistic constraint counting algorithm running on weighted interaction networks. This algorithm can be combined with the methodology of making a dilution plot [53], where rigidity parameters are calculated as a function of energy cut-off values, instead of analyzing only a single static configuration of the protein with a fixed energy cut-off value.

At the end of this subsection we have to mention that there are plenty of algorithms performing a partial structural rigidity analysis, e.g. detecting only a dominant hinge or hinges. Part of them uses additional information from comparison and alignment of similar protein structures (see Table I in [55] for an excellent summary). Other computer programs assess rigid cluster decomposition, but without complete accuracy, for example RIBFIND [56] performing neighbourhood based clustering, or applying rigid cluster decomposition based on local flexibility properties [57]. Modelling protein flexibility docking simulations is also crucial for drug design (see Tables II and III in [55]).

**Local flexibility measures of protein structures**
The most widely used local flexibility index is the temperature B-factor (Debye–Waller factor) determined by X-ray crystallography [58]. The atomic B-factor describes the mean square displacement of an atom. An averaged B-factor value belonging to the αC atom can be used to describe the mobility of a residue. A very detailed study on B-factors is done by Bhalla *et al.* [59], who studied B-factors for all secondary structure elements separately. Their results are organized by organisms, molecular functions and cellular localization. In general, the mobility of atoms or residues is not isotropic, but anisotropic. Anisotropic B-factors having 6 parameters can be measured only with resolution typically better than 1.5 Å [60]. A more generalized description, the so-called TLS (translation/libration/screw) model [61, 62], involves description of statistical translational, rotational displacements and screw motions (involving 20 parameters). It is well known that crystal contacts significantly reduce atomic fluctuations [63]. Moreover, the measured B-factors also contain the effect of lattice defects and lattice vibrations [60]. Therefore, flexibility indices based on dynamic models are also very important, especially for native states. Since there are segments in many proteins, which move on a much larger scale than the rest of the protein, the normalization of B-factors is important and can be helped by structural rigidity analysis [64]. Correlation coefficients between NMR and X-ray measurements and four different kinds of models are summarized by Livesay *et al.* [65].

Beyond the mean square displacement, other physical quantities are also useful in describing local protein flexibility. It was shown that both hydrophobicity and distance from protein surface play an important role in determining residue mobility [66]. Amino acid volume is inversely proportional to the atomic B-factor [67]. The hydrophobicity–volume product gives an even better description [68]. Number of close contacts also gives a very simple and good description of local flexibility [69]. All the measurements, models, and quantities which can measure local flexibility of proteins are summarized in Table 1 [38, 59, 60, 62, 65, 67–80]. It is important to note that the number of rotating bonds of amino acids and their side-chain flexibility properties are not equal to each other. Flexibility/mobility order of amino acids was described in several publications [81, 82].

**Table 1:** Local flexibility plots of proteins in the literature

| Measurements | | |
|---|---|---|
| **Method** | **Quantity** | **References** |
| X-ray crystallography | temperature B-factor | [59, 68, 70] |
| NMR spectroscopy | order parameter | [69, 71] |
| neutron scattering | force constant | [72] |

| Correlating quantities | |
|---|---|
| **Quantity** | **References** |
| volume | [67] |
| hydrophobicity-volume product | [68] |
| contact number/density | [67, 69, 73, 74] |
| disorder probability | [75] |

| Calculated mean square values from protein models | |
|---|---|
| **Model** | **References** |
| GNM (Gaussian network model) | [76, 77] |
| ENM (elastic network model) | [73, 78] |
| TLS | [60, 62, 79] |
| FIRST (flexibility index) | [38] |
| DCM (distance constraint model) | [65] |
| Brownian dynamics (force constant) | [80] |

**Rigidity and flexibility in protein function**

There are several interesting results and hypotheses describing a possible role of rigidity and flexibility in protein function. Both theory and measurements show that active site residues tend to be locally less flexible/mobile than others [42, 70]. Moreover, active sites usually occur in global hinge centres indicating a low mobility [83]. Analysis of protein structure networks and elastic network models indicate that active centres usually have non-redundant, unique connections, and often behave as 'discrete breathers' displaying a unique mobility pattern as compared to the rest of the protein. Discrete breathers occur at the stiffest regions of proteins, and may display a long-range energy transfer [84–87]. On the contrary, increased flexibility of some activation segments may contribute to inactive, zymogen structures of protease enzymes [88], and differential flexibilities of the activation domains may also govern substrate-specific catalysis in the trypsin/chymotrypsin family of proteases [89].

The mechanism of allosteric changes attained a lot of attention in the last century [85, 90–95], but its major molecular mechanisms are still not completely understood. Current molecular mechanisms include the propagation of a 'frustration front' [96], where frustration means that certain residues have accumulated a tension, which prompts their unusually large dislocation, once the allosteric ligand triggered the release of this tension. A recent study [41] on 51 pairs of proteins in active and inactive state using the FIRST algorithm [38] found that rigid paths connect the effector and catalytic sites in 69% of the data set. An increased rigidity in active state relative to the inactive state has also been observed. However, 31% of the studied structures failed to follow this trend [41]. Currently we do not have a coherent picture of the involvement of flexibility/rigidity changes in protein function, which prompts further studies of this exciting field.

**Rigidity as a stability factor in thermophilic proteins**
Proteins from thermophilic organisms usually show high thermostability. Plenty of structural parameters contribute to thermostability including rigidity [39], but there is no unique factor determining thermostability [97]. A remarkable study of Radestock and Gohlke [98] using the FIRST algorithm [38] analyzed not only the static structural rigidity, but simulated increasing temperatures by breaking the weakest interactions as the temperature increased. From the dilution plot a critical temperature was determined, where the rigid cluster of the protein structure became fragmented. It is important to note, that this study can only be performed on proteins, which have a large rigid backbone. A similar phase transition study [99] was accomplished for 12 pairs of proteins using the CFinder clustering algorithm [100] looking for changes in densely connected k-cliques in protein structure networks. Additional properties of thermophilic proteins may also increase their rigidity such as the existence of aromatic clusters, which were indeed found to be relatively rigid regions [101]. Rigidity analysis remains an important tool to understand the structure of thermophilic proteins. However, due to the complexity of interactions, the inclusion of a neural network model may be useful as exemplified by on the comparative study of mesophilic and thermophilic proteins using amino acid sequences [102].

## RIGIDITY AND FLEXIBILITY OF THE CYTOSKELETAL NETWORK
The combinatorial rigidity analysis is applicable for protein complexes, where the 3-dimensional structure or at least the interaction network is known. For larger protein structures, such as polymer networks, where the filaments and the network structure are formed statistically, statistical models may be used. The cytoskeletal network is a highly complex polymer network containing actin, tubulin and intermediate filaments and their complex connection structures.

When modelling a single polymer filament the rotational freedoms along its length are not strictly free, because bending requires energy. Therefore instead of a freely joined chain model (analogous to a bar-joint model), the continuous worm-like chain (or Kratky-Porod) model [103, 104] is more adequate. In the worm-like chain model the rigidity/flexibility property of the filaments is described by the so-called persistence length. This is the correlation length of the tangent vector along the polymer path (proportional to the average curvature and depending on the temperature [105]). Experimental values for the persistence length of one-dimensional biological filaments vary from 2 nanometers for the giant protein, titin [106] through 50 nanometers for double-stranded DNA, 10 microns for actin filaments up to the mm-range for microtubules.

When persistence length is comparable to other length scales appearing in the polymer network, such as the total contour length of polymers, or distance of crosslinks or branch points, the so-called semi-flexible polymer network model has to be applied. This is the case for actin filament network in the cytoskeleton [107, 108], but similar semi-flexible networks can be formed by segments of DNA [109], of aggregated amyloid fibrils [110], or of self-assembled peptide nanotubes [111]. A semi-flexible polymer network shows more complex properties than conventional ones. This improved complexity seems to be needed for the regulation of the cytoskeletal network [112]. On the one hand a striated muscle cell has an extremely structured actin cytoskeleton. On the other hand motile cells, like leukocytes or fibroblasts, have a very dynamic cytoskeleton. A semi-flexible network phase space with four expected regions has been found for these latter, more complex, dynamic systems having an affine entropic, an affine mechanical, a non-affine and a rigid region (see Figure 6 in [113], or Figure 3 in [114]). The cytoskeletal F-actin filaments seem to be at the cross-over regime between the non-affine and affine entropic regimes, therefore the network may shift from a large linear-response of the non-affine regime to the strain-hardened affine entropic regime.

The cytoskeletal network ensures the mechanical resistance and shape of the cell [115], but it should also give an opportunity to change cellular shape, which is critical in cellular movements. To accomplish this complex task the cytoskeletal network forms a tensegrity structure [116]. Microtubules constitute a compression structure and actomyosin network itself can act like a tensegrity structure. Actomyosin tensegrity structure is formed once a global balance between local contraction and neighbouring bond stretching is achieved. This tensegrity structure provides interplay between local force generation and collective action [117]. Moreover, tensegrity structures are organized in a hierarchical manner having a core, giving the opportunity to communicate with the nucleus of the cell through the intermediate filaments. The nucleus itself also has its own tensegrity system [116]. The hierarchy of tensegrities are present both in lower scale, where the viscoelastic properties of stress fibres can be modelled as a multi-modular tensegrity structure [118] and in higher scales spanning from tissues up to the skeleton, where bones represent the compression elements, while muscles, tendons and ligaments represent the tension elements of the tensegrity structure [116].

## FUNCTIONAL FLEXIBILITY AND RIGIDITY OF COMPLEX BIOLOGICAL NETWORKS

The complexity and hierarchical nestedness of biological networks infers several specific properties of these networks at the systems level. Most biological networks exert their functions as a part of a higher level biological system. Therefore the interaction of these networks with their environment has a profound effect on their structure. As a sign of this, biological systems often show a bimodal character being either rigid/stable/robust or flexible/adaptive according to their actual environment, and the dynamics of environmental changes. Many biological networks display a cyclic dynamics, where the above two states (and others) are changed periodically. However, most biological systems have to be rigid (stable) and flexible at the very same time. Therefore, biological systems had to develop a highly complex regulatory mechanism to set the proper state of rigid or flexible states [4].

Biological networks usually develop a modular structure. Modules help the separation of functional units and their interaction offers an easy, but complex way to regulate the stability (rigidity) and adaptability (flexibility) of the system [119]. Modules also participate in building the hierarchical nested structure of biological systems. Extreme modular structures (such as almost completely separated, very rigid modules, or over-confluent, too flexible modules) cause extreme behaviours leading to un-adaptive or hyper-adaptive systems, with extremely large or negligible 'memory', respectively. An optimal balance ensures both the capability for adaptation and the preservation of beneficial adaptive changes (Figure 4 [120]).

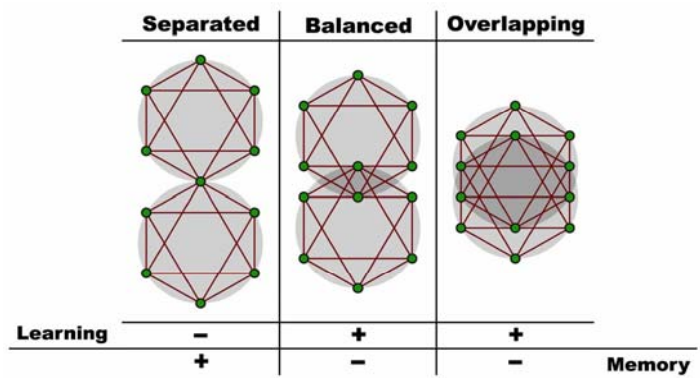

**Figure 4:** Bimodal behaviour of modular complex systems. Structure of complex systems often shows a bimodal behaviour, where the system is either composed of highly coherent, rather rigid communities having a low overlap resembling of a cumulus cloud (see left side of figure, 'Separated') or has rather flexible communities, having a high overlap resembling to a stratus cloud (right side of figure, 'Overlapping'). Extremely rigid systems have difficulties to adapt, to 'learn'. However, once they have changed, they preserve the change (they have 'memory'). On the contrary, extremely flexible systems may change easily (they 'learn' well). However, these systems cannot preserve the change: they have no 'memory'. The optimal – and most complex – solution is a simultaneous development of network flexibility and rigidity (middle panel, 'Balanced') [120].

Structural rigidity is defined for networks embedded in a metric space. Most biological networks discussed in this section are conceptual networks lacking a real metric space in their connection structure. However, using the concept that flexibility describes the internal degrees of freedom, structural rigidity/flexibility and functional rigidity/flexibility may be connected. The degrees of freedom in a functional response of a living system are represented in its microstructure. Giving an example to this relationship let us think about learning process in a neuronal network. If the structure of the neuronal network remains unchanged, because it has no internal freedom, its response to an external influence remains the same, i.e. its behaviour is repeating itself, and said to be rigid. Despite of these emerging conceptual links, the delineation of relationships between the structural properties and the dynamical behaviour of a complex network remains a very big challenge.

**Functional rigidity of protein-protein interaction networks**

Using the in-house developed ModuLand program package [121, 122] to identify extensively overlapping network modules it was recently showed that yeast protein-protein interaction networks displays a stratus to cumulus type of change when responding to a large variety of stresses [119]. In this change the interactome modules became more condensed after heat shock, and the number of inter-modular links decreased. The partial disappearance of inter-modular connections proved to be useful, since it a.) spared links and thus energy; b.) slowed down the propagation of damage; c.) allowed a larger independence of modules and thus a larger exploration radius of their adaptive responses. Importantly, residual bridges between modules were often composed by proteins playing a key role in cell survival [119].

**Functional rigidity of metabolic networks**
Metabolic networks are directed networks based on metabolic pathways. It is possible to express all possible steady-state flux distributions of the network as a nonnegative linear combination of the so-called extreme pathways. These conically independent extreme pathways are edges of the steady-state solution cone in the high dimensional flux space, were each axis corresponds to a directed edge (flux) in the network [123]. This method is appropriate to compare pathway redundancy between different metabolic networks [124, 125], therefore gives a possible quantification for the rigidity properties of the network. Importantly, the bimodal distribution of rigid and flexible network structures [4, 120] mentioned in the introduction of this chapter is also displayed by metabolic networks [122]. In their seminal review Bateson *et al.* [126] described small and large phenotypes of human metabolism resembling to the rigid, cumulus-like and to the flexible, stratus-like systems, respectively.

In case of overproduction metabolic networks have to provide a flexible redirection of fluxes. This enables the definition of metabolic rigidity/flexibility which was reviewed by Stephanopoulos *et al.* [127]. Nodes involved in dynamic flux re-partitioning to meet metabolic demands were called as flexible nodes. Metabolites of flexible nodes had similar affinities for their enzymes (representing edges of the network) and reaction velocities were also of similar magnitude. The flux through each branch was controlled by feedback inhibition by the corresponding terminal metabolite. A node was said to be strongly rigid, if the split ratio of one or more of its branches was tightly controlled. This was commonly achieved by a combination of feedback control and enzyme trans-activation by a metabolite in an opposite branch [127].

**Functional rigidity of gene regulatory networks**
Interactions among genes have a key role in the realization of any phenotype. Flexible relationships among nodes are a major source of robustness against mutations in genetic networks [128], and cause the emergence of new properties [129, 130]. The results of Swinderen and Greenspan [131] imply a considerable flexibility in *Drosophila* gene interaction network. In yeast gene regulatory networks hubs were found both as repressed and repressors, the intra-modular dynamics were either strongly activating or repressing, whereas inter-modular couplings remained weak. The main contribution of the repressed hubs was to increase system stability, while higher order dynamic effects (e.g. module dynamics) mainly increased system flexibility. Altogether, the presence of hubs, motifs and modules induced few flexible modes sensitizing the network to external signals [132, 133].

**Functional rigidity of neuronal networks**

A neuronal network is a set of interconnected neural cells. The most important segment of the neuronal network is the 'active sub-network' of the simultaneously active neural cells. Changes in activity patterns provide an extreme flexibility at this network level. Learning and memory formation play a key role in the remodelling of neuronal networks in both their 'active sub-network' and persistent structure level [134]. Parts of the neuronal system have to be extremely stable, like the neuronal network that controls breathing [135]. But in general, neuronal systems must show a highly complex multitude of simultaneous stability and flexibility. In conjunction to this, brain functions have to be noise tolerant to a certain level, which may be ensured by a certain level of rigidity in the underlying network. However, dynamic flexibility after a certain threshold becomes crucial to respond to emergency situations. Therefore the function of cortical networks is directly linked to the interplay between stability/rigidity and flexibility. This ability to switch between states or tasks is a key point of flexible behaviour [136]. Given a fix amount of resources there is always a necessary balance between resources devoted to the current task and resources devoted to monitoring non necessary relevant information. This is a balance between accuracy and quickness of the current task, and the ability to switch to a different task, which has the so-called switch cost [137]. Computer simulations showed that a regulated complex neurodynamics, which can shift its balance between sensitivity and stability, can result in an efficient information processing [17]. Moreover, a recent excellent study uncovered that flexibility, defined as the change in the association of a node to different modules of the human brain, predicts the capability of learning [138]. The inclusion of complex network features assuring the delicate balance between rigidity and flexibility will become an important topic in the further development of artificial neural networks solving even more complex learning tasks than those helped by these systems today.

**Key Points**

- An appropriate balance between rigidity and flexibility is crucial for proper functioning of complex biological systems.
- Structural rigidity and functional rigidity lay often rather close to each other in our everyday thinking but were approached by different methodologies so far.
- Combinatorial graph theory proved to be a useful method in the characterization of structural rigidity/flexibility. However, the use of information and matroid theories may provide novel generalizations of these key properties.
- Dimension-independent generalizations of structural rigidity on weighted networks will be essential for future progress.
- Using the concept that flexibility describes the internal degrees of freedom, structural rigidity/flexibility and functional rigidity/flexibility may be better connected in the future. However, currently this poses a large challenge for us in the field.

**FUNDING**

This work was supported by research grants from the Hungarian National Science Foundation [OTKA K83314] and by the Hungarian National Research Initiative [TÁMOP-4.2.2/B-10/1-2010-0013].

**Acknowledgments**

Authors would like to thank members of the LINK-group (www.linkgroup.hu), especially András Szilágyi for helpful comments.